**AlScN: A III-V semiconductor based ferroelectric**

*Simon Fichtner\*, Niklas Wolff, Fabian Lofink, Lorenz Kienle, Bernhard Wagner*

S. Fichtner, Prof. Dr. B. Wagner
Materials and Processes for Micro/Nanosystem Technologies
University of Kiel, Institute for Material Science
Kaiserstr. 2, 24143 Kiel, Germany
E-mail:sif@tf.uni-kiel.de

N. Wolff, Prof. Dr. L. Kienle
Synthesis and Real Structure
University of Kiel, Institute for Material Science
Kaiserstr. 2, 24143 Kiel, Germany

S. Fichtner, Dr. F. Lofink, Prof. Dr. B. Wagner
Fraunhofer Institute for Silicon Technology (ISIT),
Fraunhoferstr. 1, 25524 Itzehoe, Germany

*Ferroelectric switching is unambiguously demonstrated for the first time in a III-V semiconductor based material: $Al_{1-x}Sc_xN$ – A discovery which could help to satisfy the urgent demand for thin film ferroelectrics with high performance and good technological compatibility with generic semiconductor technology which arises from a multitude of memory, micro/nano-actuator and emerging applications based on controlling electrical polarization. The appearance of ferroelectricity in $Al_{1-x}Sc_xN$ can be related to the continuous distortion of the original wurtzite-type crystal structure towards a layered-hexagonal structure with increasing Sc content and tensile strain, which is expected to be extendable to other III-nitride based solid solutions. Coercive fields which are systematically adjustable by more than 3 MV/cm, high remnant polarizations in excess of 100 µC/cm² which constitute the first experimental estimate of the previously inaccessible spontaneous polarization in a III-nitride based material, an almost ideally square-like hysteresis resulting in excellent piezoelectric linearity over a wide strain interval from -0.3% to +0.4% as well as a*

*paraelectric transition temperature in excess of 600°C are confirmed. This intriguing combination of properties is to our knowledge as of now unprecedented in the field of polycrystalline ferroelectric thin films and promises to significantly advance the commencing integration of ferroelectric functionality to micro- and nanotechnology, while at the same time providing substantial insight to one of the central open questions of the III-nitride semiconductors – that of their actual spontaneous polarization.*

Ferroelectrics possess a unit-cell originating spontaneous electrical polarization with a spatial orientation that can be altered from one stable state to another under an applied electric field. This makes them a distinct class with increased functionality among the piezoelectric materials. The drive towards miniaturization of piezoelectric sensors and actuators (microelectromechanical systems, MEMS), the introduction of ferroelectric functionality into integrated circuit (IC-) technology as well as numerous emerging applications based on polarization control have led to substantial scientific and commercial interest in ferroelectric thin films. [1–5] Many of the more important ferroelectrics are perovskite oxides, with typical disadvantages such as low paraelectric transition temperatures, non-linear displacements or limited compatibility with e.g. complementary metal-oxide-semiconductor (CMOS) or III-nitride technology – issues which so far impede the universal availability of ferroelectric functionality in microtechnology. [2,6]

In the wurtzite-type structure (space group $P6_3mc$), the III-V semiconductors AlN, GaN and InN possess a spontaneous polarization along their *c*-axis, which originates in the separation of the group-III and nitrogen atoms in individual planes. [7] Therefore, two antiparallel polarization directions exist: N-polar and metal-polar (e.g. Ga- or Al-polar) (**Figure 1** (a)). The pure wurtzite-type III-nitrides are thus pyroelectric materials, but not ferroelectric, as it is accepted that their polarization direction cannot be switched with electric fields below their individual dielectric breakdown limit. [2] Due to the higher piezoelectric coefficients of AlN compared to GaN or InN, AlN is generally preferred for piezoelectric applications. [7,8] Akiyama *et al.* demonstrated that the piezoelectric response of solid solutions constituted from AlN and ScN increases monotonously with Sc content, as long as the wurtzite structure is maintained. [9,10] This can be related to the existence of a metastable layered-hexagonal phase in ScN, [11,12] which in turn flattens the ionic potential energy landscape of e.g. wurtzite-type $Al_{1-x}Sc_xN$. Consequently, the wurtzite basal plane as well as the internal parameter $u$ (the length of the metal-nitrogen bond parallel to the *c*-axis relative to the lattice parameter *c*) increases, i.e. the layered hexagonal phase is approached, particularly at the Sc sites. [11,13] While the layered hexagonal structure itself is non-polar, it can be seen as a transition state ($u = ½$) between the two polarization orientations of the wurtzite-structure (Figure 1 (a)). The sign of the polarization switches once $u$ passes ½. Due to the flattening of the ionic potential towards the hexagonal phase, the energy barrier which is associated with $u = ½$ is set to decrease as the Sc content is increased. Strain engineering of the wurtzite basal plane should enable a further decrease of this barrier, as was predicted by Zhang *et al.* for $Ga_{0.625}Sc_{0.375}N$. [13] In the same case, the electric field necessary to achieve ferroelectric polarization switching was also calculated to be potentially below the dielectric breakdown limit of pure GaN. [13] Aside from the incorporation of ScN into AlN, GaN or InN, other metal nitrides such as YN or MgN-NbN were predicted and/or found to lead to a similar softening, increased *a*-lattice parameter or improved piezoelectric response. [12–16] The key to ferroelectric switching however remains that the energy barrier between the two polarization states of the wurtzite structure can be lowered sufficiently – either by increasing the ratio of the non-III metal or via strain engineering, while to some extent preserving the dielectric breakdown resistance of the pure III-nitride. In spite of this theoretical motivation, no experimental evidence of ferroelectric III-nitride semiconductors has been reported to date.

**Experimental Details**

Polycrystalline $Al_{1-x}Sc_xN$ films were prepared by reactive sputter deposition on oxidized 200 mm (100) Si wafers covered with an AlN/Pt bottom electrode. The relevant process parameters for the $Al_{1-x}Sc_xN$ films deposited from dual targets (all except $Al_{0.64}Sc_{0.36}N$) were previously published. [17] $Al_{0.64}Sc_{0.36}N$ was deposited from a single alloy AlSc target with a nominal Sc content of 43 at.% and a purity of 99.9 at.%. Here, the DC power was set to 600 W, the gas flows into the chamber to 7.5 sccm of Ar and 15 sccm of $N_2$, while the substrate was kept at 400°C during deposition. The film thickness was set to either 400 nm (all films with $x = 0.27$), 600 nm ($x = 0.32; 0.36; 0.40$) or 1 μm ($x = 0.43$). The PZT film with a thickness of 600 nm was derived via sol-gel deposition with previously published process details. [18] (Piezo-) Electrical characterization took place on parallel plate capacitors with Pt top electrodes structured by lift-off. A commercially available aixACCT double beam laser interferometer (*P-E* loops, inverse piezoelectric effect, square 0.25 mm² top electrodes for *P-E* Loops, square 1 mm² top electrode for strain measurements) and 4-point bending probe (direct piezoelectric effect, 16 mm² top electrodes) was used for the (piezo-) electric characterization. [19,20] P-E loops were measured with a triangular voltage input at 711 Hz and 411 Hz (for correction only, see supporting information). The inverse piezoelectric effect was measured with a triangular voltage input at 211 Hz. Due to systematic errors arising from non-neglectable substrate deformation during measurements of the inverse piezoelectric effect, [21] a correction factor of 0.85 was multiplied with the measured strain response (see supporting information). Polarization inversion was initiated through a unipolar 0.1 Hz sine wave with a peak value of -200 V over 60s. The residual stress was calculated via Stoney's equation based on capacitive measurements (E+H MX 203) and profilometer line scans (Ambios XP2) to extract the substrate curvature on wafer level. [22] Temperature treatments were performed in consecutive steps of 5 minutes each in ambient atmosphere. Etching experiments to resolve the polarity distribution in the $Al_{1-x}Sc_xN$ films were carried out in 85% phosphoric acid ($H_3PO_4$) and 25% potassium hydroxide (KOH) aqueous solutions at 80°C. To remove the top electrodes after ex-situ polarization inversion, ion beam etching (IBE, Oxford Instruments Ionfab 300) was employed. For the characterization of the micro- and nanostructure transmission electron microscopy was performed on three different microscopes during this investigation: Dark-field imaging was performed on a JEOL JEM-2100 (200 kV, $LaB_6$ cathode) while high-resolution TEM imaging was performed on a Tecnai F30 STwin microscope (300 kV, field emission cathode, $C_S$=1.2 mm) and a Philips CM 30 ST (300 kV, $LaB_6$ cathode, CS=1.15 mm). The cross-section sample was prepared by focused ion beam (FIB) milling using a standard lift-out method with a FEI Helios Nanolab system. The Sc content was measured by scanning electron microscopy energy dispersive x-ray spectroscopy (SEM-EDX) (Oxford x-act, 10 kV).

**Ferroelectric Properties: Polarization Hysteresis, Direct and Indirect Piezoelectric Effect**

Depending on the Sc content and the uniaxial mechanical stress of the $Al_{1-x}Sc_xN$ films, it was possible to demonstrate not only that ferroelectric switching can indeed be achieved in wurtzite-type III-nitride based solid solutions – but also that the material has exceptional properties of relevance to the core-applications of ferroelectric thin films. The studied $Al_{1-x}Sc_xN$ layers generally exhibited good *c*-axis orientation normal to the substrate, [17] although grains with secondary orientations/phases were observed for Sc contents with $x \geq 0.4$ (see supporting information).

Distinct ferroelectric polarization inversion was measured starting at Sc contents of $x = 0.27$. Below (at $x = 0.22$), dielectric breakdown occurred before reaching the coercive field $E_C$. P-E (polarization over electric field) hysteresis loops of $Al_{1-x}Sc_xN$ with $x = 0.27, 0.32, 0.36, 0.40$ and $0.43$ are given in Figure 1 (a).

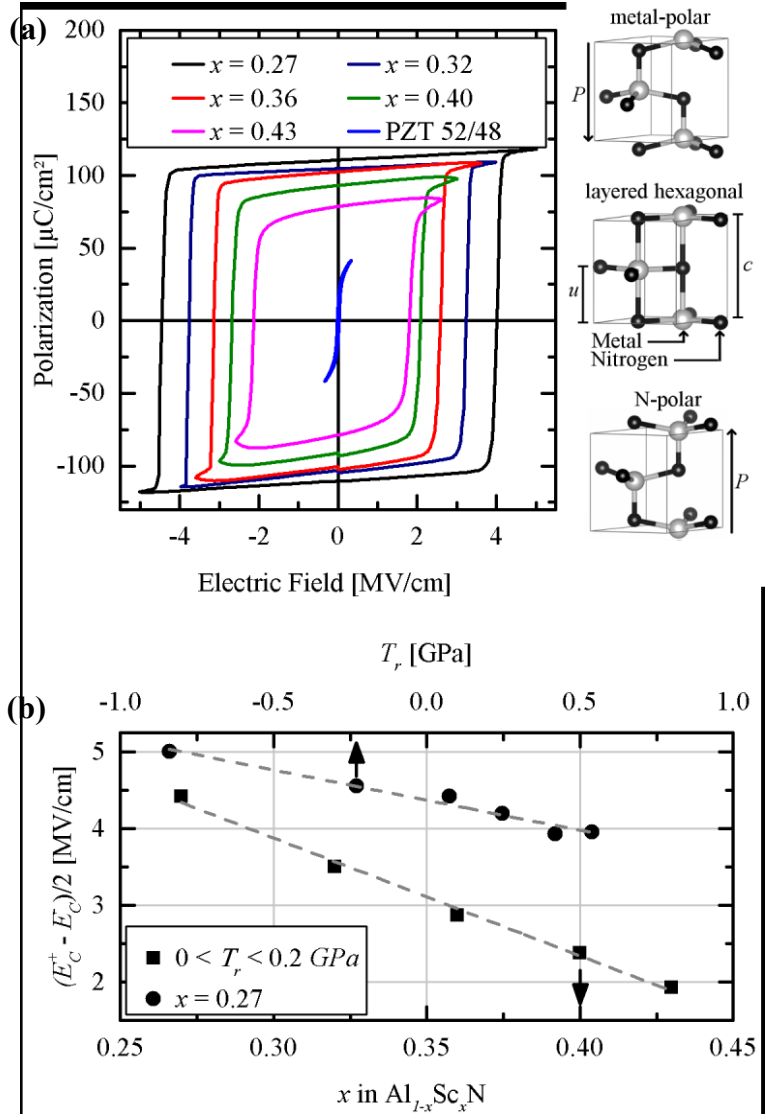

**Figure 1: (a)** P-E loops of ferroelectric $Al_{1-x}Sc_xN$ with Sc contents of $x = 0.27, 0.32, 0.36, 0.40$ and $0.43$ as well as of PZT 52/48. To the right, the structures associated with the respective polarization states are displayed. **(b)** Dependence of the mean coercive field on the residual stress $T_r$ and the Sc content of the $Al_{1-x}Sc_xN$ films. Sc content and $T_r$ were varied independently.

For comparison, the P-E loop of a $PbZr_{0.52}Ti_{0.48}O_3$ (PZT 52/48) film measured with the same parameters is displayed as well. To (at least partially) compensate the *P-E* loops with respect to the non-negligible leakage currents at higher electric fields, a modified dynamic leakage current compensation was employed (see supporting information). [23] In general, we observed very large coercive fields (beyond 4 MV/cm at $x$ = 0.27), almost undiminished polarization between the coercive fields and high remnant polarizations (110 µC/cm² at $x$ = 0.27). The latter is significantly above theoretical predictions of the spontaneous polarization in both pure AlN ($\approx$ 10 µC/cm²) and Al$_{1-x}$Sc$_x$N ($\approx$ 30 µC/cm² at $x$ = 0.5) which were made using the zincblende structure as the reference [7,24,25]. The polarization, and its trend with respect to $x$ (which in turn increases $u$ and leads to approaching the non-polar layered hexagonal structure) are however in line with a more recent prediction made in reference to the layered hexagonal structure by Dreyer *et al.* [8]. As such, it is a welcome side-effect of ferroelectricity in Al$_{1-x}$A$_x$N that the previously experimentally virtually inaccessible spontaneous polarization of AlN can therefore projected to be indeed above 100 µC/cm², rather than below 10 µC/cm² - thereby providing an answer with regard to the uncertainty associated with the polarization constant of the III-nitrides and confirming the approach by Dreyer *et al.*.

The almost ideal box-like shape of the polarization hysteresis and the large coercive fields can be related to a still sizeable energy barrier associated with the hexagonal phase, good compositional homogeneity and the wurtzite structure itself, which allows only 180° domain rotations. In terms of shape and polarization magnitude, the *P-E* loops of Al$_{1-x}$Sc$_x$N remind of certain measurements on epitaxial ferroelectric thin films [26–28], albeit without the need for a specific template to facilitate epitaxy, and therefore with increased compatibility and ease of fabrication.

The gradual lowering of the switching barrier ($u$ = ½) with increasing Sc content results in a linear decline of the coercive field (Fig. 1 (b)), from above 4 MV/cm (Al$_{0.73}$Sc$_{0.27}$N) to less than 2 MV/cm (Al$_{0.57}$Sc$_{0.43}$N). Just as the Sc content distorts the wurtzite-type crystal structure by expanding its basal plane and increasing $u$, lateral mechanical straining of the films can be used to same end. Similar to what has been reported for pure AlN, permanent lateral mechanical stress of a well-defined magnitude was induced to the Al$_{1-x}$Sc$_x$N films by varying the Ar partial pressure of the sputter gas. [29,30] Adjusting the mechanical stress $T_r$ in $Al_{0.73}Sc_{0.27}N$ films from about -0.8 GPa to +0.5 GPa resulted in a linear decline of the coercive field by more than 1 MV/cm. Therefore, a high degree of flexibility exists to systematically adjust the switching voltage towards a value favoured for the intended application by independent variation of the residual stress or the Sc content. The range of achievable switching voltages is further extended by the observation that polycrystalline Al(Sc)N can possess a preferential c-axis orientation already on the first 10 nm from the substrate (see supporting information) and functional films of corresponding thicknesses are therefore feasible. [31] Thus, both high switching voltages (> 100 V) of advantage to linear piezoelectric excitation as well as low switching voltages in ranges of relevance for memory applications (< 10 V) could be realized.

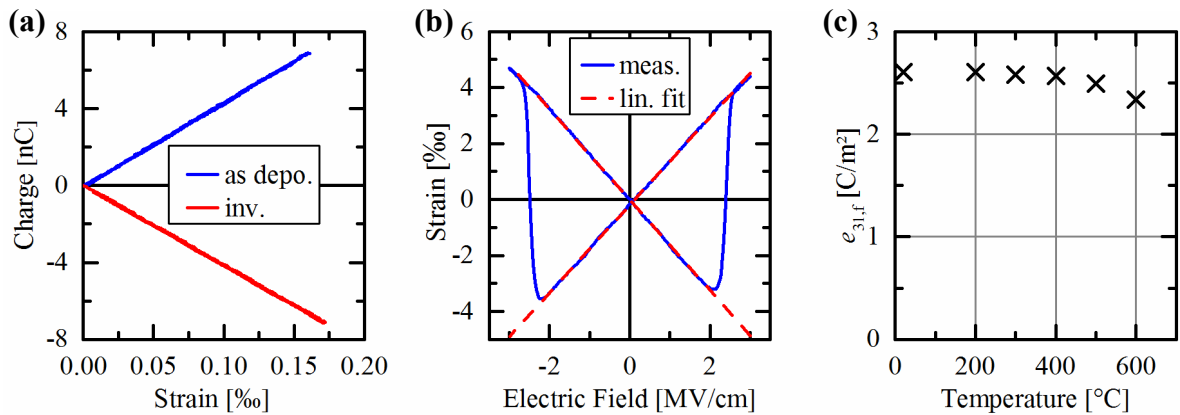

**Figure 2: (a)** Direct piezoelectric effect: Charge-strain curves of $Al_{0.64}Sc_{0.36}N$ as deposited and after ferroelectric polarization inversion. **(b)** Converse piezoelectric effect: Longitudinal strain response of $Al_{0.64}Sc_{0.36}N$. **(c)** Transverse piezoelectric coefficient after polarization inversion and successive temperature treatments at the indicated temperatures without subsequent repolarization.

Measurements of the direct and inverse piezoelectric effect in ferroelectric $Al_{0.64}Sc_{0.36}N$ are given in **Figure 2** (a) and (b) respectively. Both directions imply the possibility for virtually complete polarization inversion during ferroelectric switching. For measurements of the direct piezoelectric effect on an inverted sample, a simple 1-minute poling procedure at room temperature was used to switch between the polarization states. Under this procedure, the effective transverse piezoelectric coefficient $e_{31,f}$ was inverted from an as-deposited -2.90

C/m² to 2.76 C/m². Both values can be considered high for an AlN-based solid solution [17,32]. Repeated measurements of the piezoelectric response up to 30 weeks after polarization inversion did not show measurable degradation.

Moreover, the polarization inversion was conserved up to at least 600°C: $e_{31,f}$ declined only slightly during the temperature treatment steps (Figure 2 (c)). Consequently, 600°C can be seen as a lower limit for the paraelectric transition temperature of $Al_{0.64}Sc_{0.36}N$. Beyond 600°C, a degradation of the electrodes prohibited further electrical characterization of the capacitors.

The longitudinal displacement butterfly curve (Figure 2 (b)) of the inverse piezoelectric effect has broad linear regimes with almost equal slopes which correspond to an effective longitudinal piezoelectric coefficient $d_{33,f}$ of 15.7 pm/V and -16.2 pm/V. Compared to state of the art polycrystalline ferroelectric thin-films, both the width of the linear strain regime of 0.7% and its symmetry around the field axis are outstanding [33].

**Evidence for Genuine Ferroelectricity Associated with the Wurtzite Crystal Structure**

It is long known that not only true ferroelectricity results in P-E hysteresis loops, but also electrets, finite conductance or p-n and Schottky junctions can lead to P-E measurements which resemble ferroelectricity [3,34–37]. In contrast to these spurious effects, which are connected with charge migration on length scales up to the film thickness, ferroelectricity is based on a stable and repeatable polarization reorientation on unit-cell level. The ultimate evidence for ferroelectric switching would therefore be the in-situ observation of the underlying atomic displacement under and after the application of an external electric field. While such experiments were performed in the past [38], the high coercive fields of AlScN would add additional challenge to an already demanding investigation. Alternatively, ex-situ polarization inversion could be used with methods were contrast specific for the unit-cell orientation can be obtained between pristine regions and switched regions. For the case of the wurtzite semiconductors GaN, AlN, InN and ZnO, a standard method which provides such unit-cell orientation specific contrast is wet-etching in both acids ($H_3PO_4$) and bases like aqueous KOH and TMAH [39–45]. While N-polar surfaces etch readily and with distinct residues, metal polar surfaces do barely etch at all and initially remain smooth, with the exception of local defects and inversion domains.

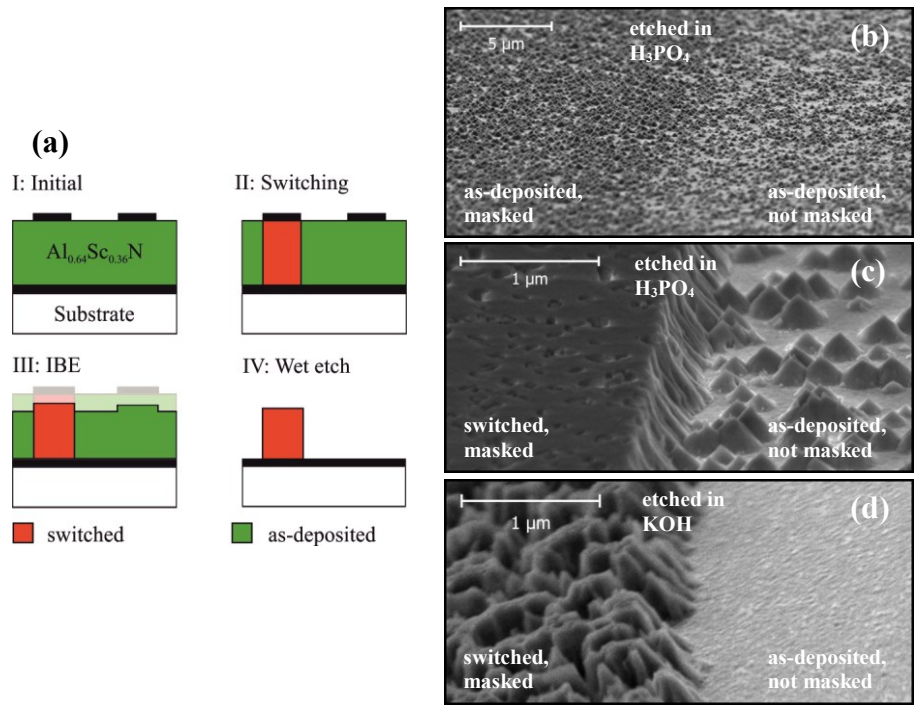

**Figure 3: (a)** Sketch of the experimental procedure to determine the polarity of $Al_{0.64}Sc_{0.36}N$ via wet etching. **(b)** SEM image of the step from a masked, as deposited, region to an unmasked as-deposited region (faintly visible as a line through the center of the image) after etching in $H_3PO_4$ for 5 min. **(c)** Step from a switched region to an as-deposited region after etching in $H_3PO_4$ for 5 min. **(d)** Step from a switched region to an as-deposited region after etching in KOH for 15 s.

To investigate whether this anisotropy can be observed in $Al_{1-x}Sc_xN$, samples from the same wafer as used for the piezoelectric characterization above had some of their capacitors switched, while others were kept as deposited. Subsequently, their top electrode was removed via IBE with an intentional 50 nm overetch into to $Al_{0.64}Sc_{0.36}N$ film **(Figure 3 (a))**. This overetch was chosen in order to, one the one hand, rule out masking due to residues of the top electrode and on the other hand to remove the interface region of the $Al_{0.64}S_{0.36}N$ film, as charge injection and ionic migration typically manifests around the electrodes [35,36]. Afterwards, the samples were etched in $H_3PO_4$ and KOH until not more than residues remained on the bottom electrode outside the capacitor areas. After wet etching, polarization inverted structures were conserved with close to their full height (ca. 500 nm), having a smooth surface intermitted by deep holes due to either defects or inversion domains (Figure 3 (b)) (the presence of the latter can be assumed due to the slightly lower piezoelectric coefficient of the samples, figure 2 (a)), which merged after longer etching times in $H_3PO_4$ or when in etching in KOH, due to generally more rapid etching in the latter case. On the contrary, the rest of the film etched readily and with the characteristic, cone like residues (Figure 3 (c)). Due to the comprehensive investigation of the same effect in pure wurtzite semiconductors, this result provides conclusive evidence that polarization switching on unit-cell level does indeed take place in $Al_{1-x}Sc_xN$ and that, therefore, the material is a genuine ferroelectric.

Additional evidence for ferroelectricity in $Al_{1-x}Sc_xN$ can be obtained from retention and frequency dependent measurements of the electrical polarization. Unlike the measured coercive field, the measured switching polarization of a true ferroelectric material should be largely independent of the measurement frequency. Due to $\Delta P \sim \int I(t) \, dt$, where $I(t)$ is the measured current during sweeping of the electric field, the extrema of the former have to be approximately proportional to frequency (or more precisely, the area under its peaks have to be). This is typically not the case for currents due to charge injection/leakage, which either stay constant or decrease for increasing frequencies [23,35]. For $Al_{1-x}Sc_xN$ however, the area associated with the switching current peaks does scale proportionally with frequency over more than two orders of magnitude and the measured switching polarization consequently is constant (see supporting information) – as one would expect of a proper ferroelectric. Further, the stability of the polarization in electrets and due to injected charges over time has the tendency to be lower than those of ferroelectrics (although there are exceptions) [34]. The retention behavior of $Al_{0.64}Sc_{0.36}N$ was therefore determined by using the voltage pulse sequence given in **figure 4** (a). In between the read pulses for the switching ($P_{sw}$) and non-switching polarization ($P_{nsw}$), the capacitor was shortened. Over $10^5$s, no polarization loss that could be attributed to polarization back switching was observed (figure 4 (b)). The fact that $P_{nsw}$ is still not exactly zero appears to be rather due to purely dielectric effects (leakage and polarization magnitude, figure 4(c)). The typical linear loss of polarization after ferroelectric switching which was observed over a logarithmic time scale in other materials [34,46] could not be observed in $Al_{1-x}Sc_xN$, although orders of magnitude longer measurements would be necessary to give a more definite understanding of the polarization retention in $Al_{1-x}Sc_xN$. It is however straight forward to motivate an exceptionally long retention time in $Al_{1-x}Sc_xN$, based not only on the current retention measurements, but also on the fact that in its parent material, AlN, polarization is considered to be permanently aligned.

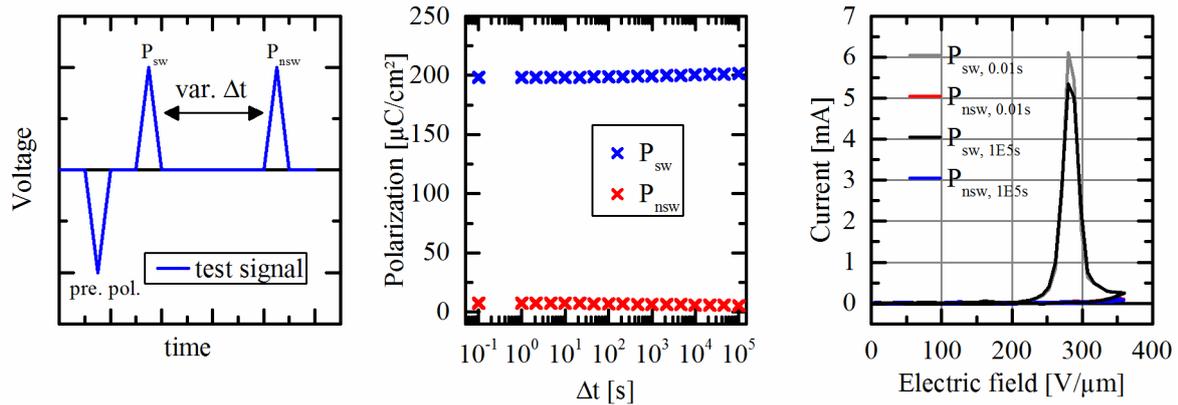

**Figure 4: (a)** Pulse sequence to determine the retention behaviour of $Al_{1-x}Sc_xN$ via switching and non-switching polarization measurements ($P_{sw}$, $P_{nsw}$) separated by a time $\Delta t$ during which the capacitor was shortened. The pulse width is 2 ms **(b)** Switching and non-switching polarization over retention time $\Delta t$.
**(c)** Initial and final current measurements, from which $P_{sw}$ and $P_{nsw}$ were extracted – a slight increase due to leakage current can be observed in the non-switching current for high electric fields, contributing to the non-zero non-switching polarization which is therefore not due to true polarization switching.

Asides from providing evidence that actual ferroelectric switching on unit-cell level does take place, transmission electron microscopy (TEM) was employed to confirm that the wurtzite structure is indeed conserved during the application of a switching field and no previously unconsidered phase is induced. In this context, an $Al_{0.57}Sc_{0.43}N$ TEM sample was prepared such that it contains two regions: one being sandwiched by Pt electrodes and subjected to ferroelectric polarization inversion and an unaltered region to allow for a structural comparison. An illustration of stitched TEM dark field images mapping out the intensity of the wurtzite-type $Al_{0.57}Sc_{0.43}N$ (0002) reflection is given in **Figure 3** (a) with the precession electron diffraction pattern (PED) of the wurtzite structure in Figure 3 (b). No significant difference in contrast between crystal columns below the Pt top electrode used for polarization inversion and crystal columns not subjected to the switching electric field was observed.

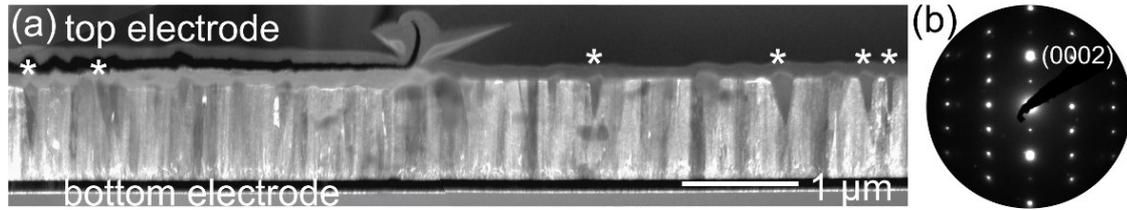

**Figure 3: (a)** TEM darkfield image mapping out the intensity of the (0002) reflection indicated in the PED pattern **(b)**. Misoriented crystals that do not exhibit the wurtzite-type (0002) texture of the matrix are displayed with a darker contrast in the brighter crystal matrix and are partially indicated by (*).

From ED experiments, the majority of the film was identified as (0002) textured $Al_{1-x}Sc_xN$, both in the pristine and the polarization-inverted region. Therefore, ferroelectricity in $Al_{1-x}Sc_xN$ can indeed be related to the wurtzite-type structure. However, owned to the high Sc content of the particular sample, a number of misoriented wurtzite grains were identified, some of which are marked by asterisks in Figure 3 (a). Moreover, small structural variations in the form of potential cubic grains and an unidentified phase were occasionally observed. Although being of high interest, the transition towards a layered hexagonal structure itself cannot be retrieved from precession electron diffraction data alone – kinematic simulations suggest that the contrast between the polar and non-polar phase can hardly be differentiated at a Sc content of $x = 0.43$. These aspects are discussed in more details within the supporting information.

To conclude, ferroelectric switching could be demonstrated in $Al_{1-x}Sc_xN$, beginning at Sc concentrations of $x = 0.27$. We expect that this material is only the first of a new group of ferroelectric wurtzite-type III-nitride based solid solutions, with likely additional candidates for this class being e.g. $Ga_{1-x}Sc_xN$, $Al_{1-x}Y_xN$ or $Al_{1-x-y}Mg_xNb_yN$. Ferroelectric switching allowed the first direct experimental observation of the switching spontaneous polarization in an AlN based material and confirmed that, contrary to most prior theoretical publications, it can reach values larger than 100 µC/cm². The unique combination of ferroelectric properties of $Al_{1-x}Sc_xN$ in concert with the excellent compatibility to existing technology platforms should make the material highly relevant for both the classical applications of ferroelectrics, like piezoelectric multilayer actuator stacks or non-volatile memory cells as well as for novel approaches based on controlling electrical polarization, e.g. in the fields of optoelectronics, multiferroic composites and III-nitride technology. [1,4,5,24]

**Acknowledgements**

The authors thank Viola Duppel from the Max Planck Institute for Solid State Research for additional TEM analyses and Prof. Bettina Lotsch for enabling these experiments. Christin Szillus is acknowledged for TEM sample preparation, Dr. Andre Piorra for the fabrication of the PZT reference and Prof. Hermann Kohlstedt for discussion and encouragement. Further, the authors thank the German Research Foundation (DFG) (CRC 1261, subprojects A3, A6 and Z1), the German Federal Ministry of Education and Research (BMBF) (grant number 16ES0632) as well as the Federal State of Schleswig-Holstein (Competence Center for Nanosystem-Technology) for funding part of this work.

**Supporting Information**

Dynamic Leakage Current Compensation

While the $Al_{1-x}Sc_xN$ films investigated in this work had typically low leakage currents (< 1 nA on 16 mm² at 1V), the current increased disproportional at higher electric fields which is attributed to the finite conductance of the films (**Figure S1**). As the coercive fields are typically within this range of disproportional increase, the corresponding P-E loops will be systematically altered by a non-dielectric current contribution, which leads to a hysteresis of non-ferroelectric origin. [37] In order to correct such distortions, a dynamic leakage current compensation (DLCC) can be used. [23] The approach is based on the assumption that the ohmic current $i_R$ through a ferroelectric film is independent of frequency, while the ferroelectric switching current $i_F$ and the current charging the capacitor $i_C$ are proportional to frequency. The total current is therefore given by:

$$i = i_R + \omega i_C^0 + \omega i_F^0, \qquad (1)$$

with $i_{F,C} = \omega i_{F,C}^0$. A compensated current $i_{comp}$ free of ohmic contributions can therefore be calculated from current measurements at two different frequencies, $\omega_1$ and $\omega_2$:

$$i_{comp}(\omega) = \frac{\omega}{\omega_1 - \omega_2}[i(\omega_1) - i(\omega_2)]. \qquad (2)$$

As in a typical current-based polarization measurement, the polarization is then given through the time-integral of $i_{comp}$. The assumption that the ferroelectric current is proportional to frequency however implies that the coercive field has to be independent of the latter. For Al$_{1-x}$Sc$_x$N, we observed a non-neglectable frequency dependency of the coercive field, which makes the application of standard DLCC unfeasible. The unusual stability of the polarization however allowed to extract the leakage current itself by assuming $i_F = 0$ during repeated unipolar voltage sweeps (which can be varied in frequency), thereby excluding any effects from the frequency dependent coercive fields. Therefore, non-switching, unipolar current measurements $i_1$ and $i_2$ were performed at two different frequencies (411 Hz and 711 Hz) and the corresponding leakage current calculated according to:

$$i_R = \frac{i_2 \omega_1 - i_1 \omega_2}{\omega_1 - \omega_2}. \qquad (3)$$

Subtracting eq. (3) from eq. (1) then results in the compensated current, which can again be integrated to provide the electrical polarization. The compensated current is thus based on 5 successive current measurements (**Figure S2**): A bipolar voltage sweep with both positive and negative switching currents (at 711 Hz in this work), two non-switching unipolar voltage sweeps for both positive and negative voltages at a first frequency (711 Hz) and two further non-switching unipolar sweeps at a second frequency (411 Hz). In addition, between two non-switching measurements, a polarization reversal sweep is performed.

The resulting compensated polarization (**Figure S3**) appears to be virtually free of leakage current induced hysteresis, although small distortions, likely due to finite conductance, remained at high electric fields.

One could argue that a similar leakage current could be obtained by simple averring between the forward and backward direction of a non-switching voltage sweep – without the need to perform measurements of the non-switching current at different frequencies. However, the so derived leakage current showed different current offsets for negative and positive voltage sweeps. The asymmetry of $i_R$ with respect to the origin then led to artificially opened hysteresis loops.

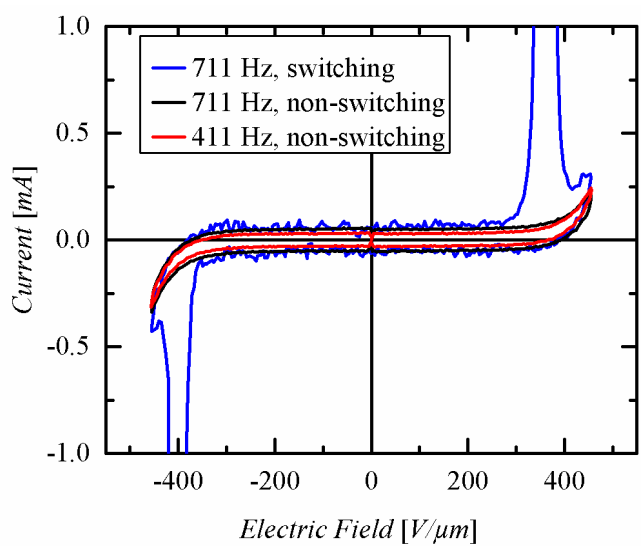

**Figure S1:** Measured switching and non-switching currents from an $Al_{0.73}Sc_{0.27}N$ sample used as basis for leakage current compensation. Each non-switching current consists of two separate measurements with intermediate polarization reversal.

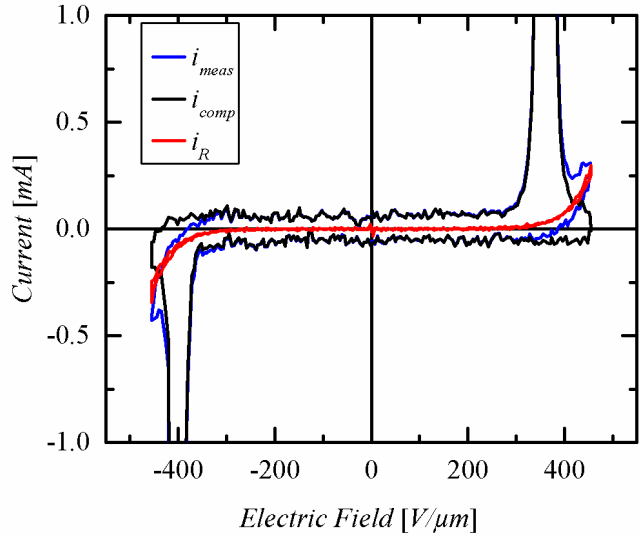

**Figure S2:** Measured current, calculated leakage current and the compensated current obtained via subtracting the leakage current from the initially measured current. The calculated leakage current is based on the measurements of Figure S1

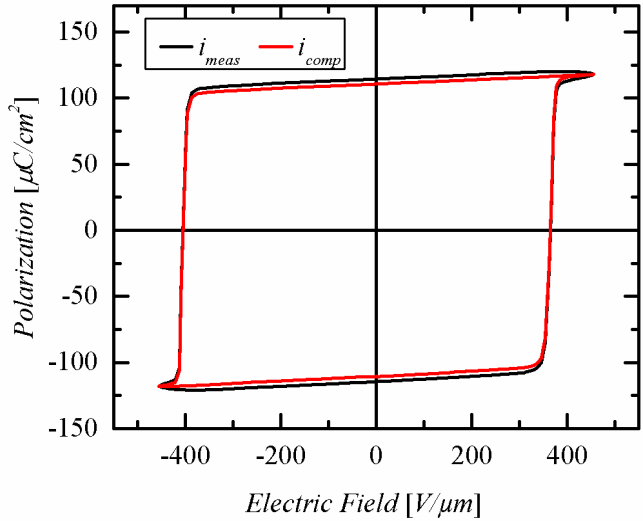

**Figure S3:** Initial and compensated (as shown in Figure 1) polarization of $Al_{0.73}Sc_{0.27}N$.

Inverse Piezoelectric Effect: Correction of the Longitudinal Strain

With $S_3$ and $E_3$ being strain and electric field normal to the capacitor surface, the longitudinal thin film piezoelectric coefficient $d_{33,f}$ is defined as

$$d_{33,f} = \frac{S_3}{E_3}, \qquad (4)$$

under the assumption of a perfectly rigid substrate. This assumption however is violated in reality, giving rise to a systematic over- or underestimation of $d_{33,f}$. Nonetheless, a quantitatively correct $d_{33,f}$ can be obtained by applying corrections determined through finite element analysis or trough measuring at a certain pad-size/substrate-thickness ratio only. [21] Sivaramkrishnan *et al.* motivated the following expression towards obtaining a quantitatively correct $d_{33,f}$ (resulting from a quantitatively correct strain measurement): [21]

$$d_{33,f} = \frac{f(r_2)d_{33,f}(r_1) - f(r_1)d_{33,f}(r_2)}{f(r_2) - f(r_1)}. \tag{5}$$

Here, $r_{1,2}$ are two ratios of electrode-size over substrate thickness and f(r) is a function of this ratio obtained by finite element modeling. In order to arrive at a correction factor for $d_{33,f}$ (and thus for $S_3$), we measured $d_{33,f}(r_1)$ and $d_{33,f}(r_2)$ of $Al_{0.64}Sc_{0.36}N$ on square electrodes with 1 mm and 0.5 mm side length using small signal measurements. On our 0.725 mm thick substrates, $r_1$ = 0.69 (0.5 mm) and $r_2$ = 1.38 (1 mm). Therefore, $f(r_1) \approx 0.83$ and $f(r_2) \approx -0.52$. [21] The longitudinal piezoelectric coefficient $d_{33,f}$ was measured to be 9.5 ± 0.2 pm/V and 17.2 ± 0.4 pm/V for the 0.25 mm² and 1 mm² electrode respectively (5 pads each). The mean corrected $d_{33,f}$ therefore is 14.22 pm/V, and $d_{33,f}/d_{33,f}(r_2) = S_3/S_3(r_2) = 0.83$.

Frequency Dependency of the Reverse Current and Polarization

**Figure S4** (a) gives the frequency dependent positive branches of the current over the electric field for eight logarithmically spaced measurement frequencies from 28 Hz to 5 kHz. The sample consisted of a 300 nm thick $Al_{0.70}Sc_{0.30}N$ film with 0.1 mm x 0.1 mm top-electrode. In order to allow comparable measurements up to 5 kHz, $E_{max}/E_c$ = 1.3 was chosen to avoid excessive contributions from leakage currents, where $E_{max}$ is the peak of the applied electrical field. The peak of the switching current itself does not scale strictly proportional to frequency (at least above 1070 Hz), but neither does the width of the peaks (figure S4 (b)) (which is plausible since ferroelectric switching is a stochastic process). Thus, the switching polarization associated with the measured currents remains constant over the whole frequency range.

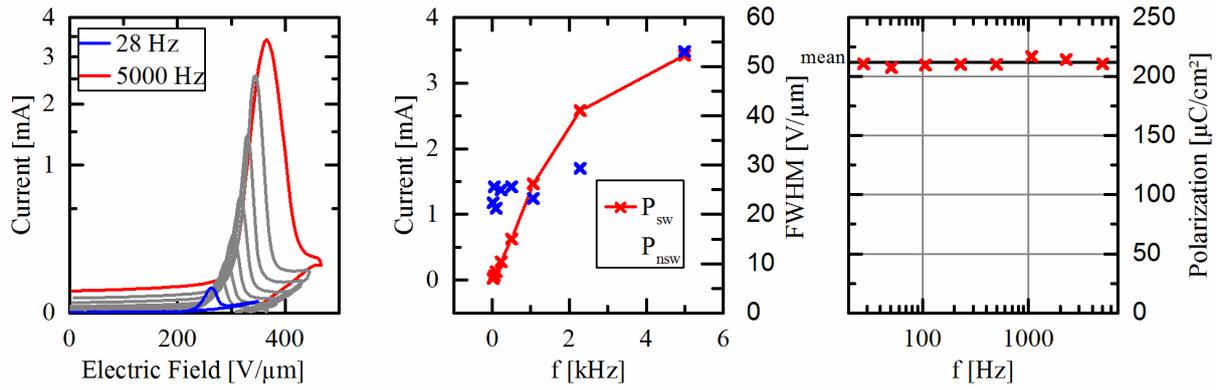

**Figure S4: (a)** Positive branch of the current hysteresis over the electric field for a 300 nm thick $Al_{0.70}Sc_{0.30}N$ film with 0.1 μm x 0.1 μm top-electrode for eight logaritmically spaced frequencies. **(b)** Current maximum and peak width from the measurements in (a). **(c)** Switching Polarization extracted from part (a).

Morphology of the $Al_{1-x}Sc_xN$ films

SEM imaging is a convenient tool in order to detect the incipient formation of secondary orientations in $Al_{1-x}Sc_xN$ thin films. These secondary orientations typically manifest in an additional elevated type of grain surrounded by a c-axis oriented matrix and can be visualized in SEM before their respective diffraction peaks appear in a standard XRD Θ-2Θ scan. [17,47] **Figure S4** therefore shows SEM-top views of films with the five Sc concentrations investigated in this work. Starting at $x = 0.40$, elevated grains associated with secondary orientations appear. The main difference in the processes used for depositing these films, compared to the ones shown in our previous work without traces of secondary orientations was a variation in Ar flow to adjust the residual stress in the $Al_{1-x}Sc_xN$ layers – which was made necessary by the correlation between residual stress and the coercive fields. [17] While our previous films had increasing compressive stress with increasing Sc content, films in this work where adjusted to be under low tensile stress ($T_r$ in [0 200 MPa]).

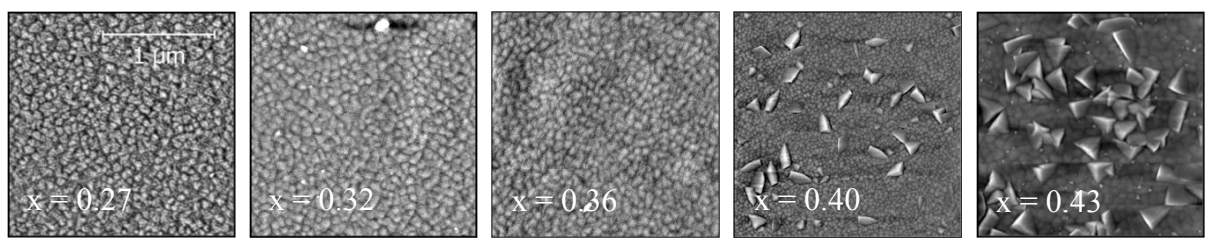

**Figure S4:** SEM surface images of the 5 films used to study the effect of Sc content on the coercive field in this work. Film thicknesses from left to right are: 400 nm, 600 nm, 600 nm, 600 nm and 1 μm.

TEM Analysis of $Al_{0.57}Sc_{0.43}N$

The wurtzite-type $Al_{0.57}Sc_{0.43}N$ film matrix in this investigation consists of highly crystalline and columnar shaped grains growing along the [0001] direction with lateral size of about 80-100 nm. Electron diffraction experiments (see **Figure S5 (a)**) indicate the systematic rotation of the single crystalline columns around their joined rotation axis [0001] following the description of the wurtzite crystal structure of AlN [48]. The c/a ratio of ≈1.44-1.46 was determined from the ED experiments and agrees very well with the observations of Akiyama *et al.* [9] for Sc rich alloys. The c-axis instability [17] of columnar shaped grains which appears with higher degree of alloying leads to a number of misoriented grains with sustained wurtzite structure. Direct information about the misorientation of the c-axis with respect to the (0002) film texture can be deduced from superposition selected-area (SA)ED pattern (Figure S5 (b), (c)) and precession (P)ED pattern (Figure S5 (d)). For visualization, each ED pattern is combined with a structural model in the respective zone axis orientation. The rather random in-plane and out-of-plane rotation of the polar axis adds a reduced contribution to the overall piezoelectric response.

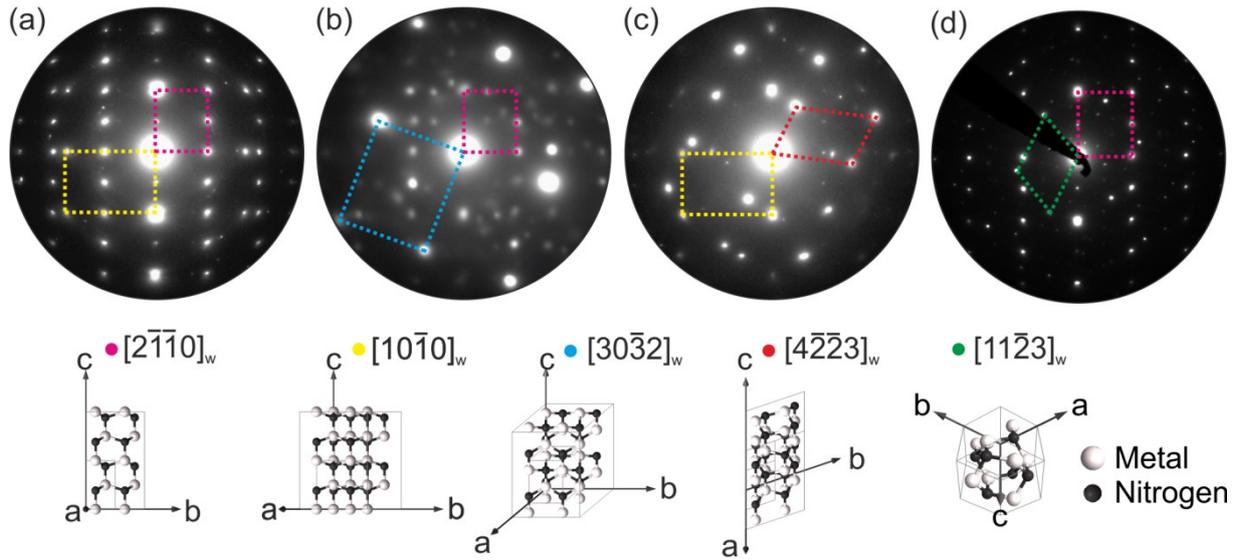

**Figure S5:** Selected area electron diffraction (SAED) pattern (a-c) and precession electron diffraction pattern (d) of the $w$-Al$_{0.57}$Sc$_{0.43}$N film with a structural representation of the observed orientations. (a) The crystal matrix is composed of columns rotated around their c-axis, showing the superposition of the $[2\bar{1}\bar{1}0]_w$ and $[10\bar{1}0]_w$ zone axis reflections. (b-d) Superposition ED pattern of misoriented crystals (blue, red, green) in the matrix as indicated.

The stoichiometry of the Sc alloy is set in the mixed phase microstructure intermediate region described by Akiyama *et al.,* reflecting the transition zone from the wurtzite structure for $x < 0.41$ to the rocksalt phase for $x > 0.46$ in Al$_{1-x}$Sc$_x$N [9,49]. Indeed this wurtzite instability is supported by high-resolution TEM images and their respecting fast Fourier transforms (FFTs) depicted in **Figure S6**. The intermittent microstructural variations from the wurtzite structure are reflected in: a) the appearance of a polycrystalline character (red dashed) close to the bottom electrode with lattice plane distances not consistent with the superimposed $[2\bar{1}\bar{1}0]_w$ pattern, but congruent with the description of Akiyama *et al.* and Höglund *et al.* for the appearance of a c-Al$_{0.60}$Sc$_{0.40}$N phase with 4.25 Å lattice parameter; b) an entire crystal with its FFT showing a typical [110] pattern of a cubic structure, however, the lattice parameter of 3.94 Å fits very well to the metastable (at ambient pressures) high-pressure phase of AlN [50], rather than to the description of rocksalt structure alloy films discussed by Höglund *et al.* [49] and Saha *et al.* [51]; c) a grain boundary between a $[2\bar{1}\bar{1}0]_w$ oriented crystal (i) and one crystal with a non-identified structure (ii), but with lattice plane distances partially coinciding with the neighbouring grain. Both crystals share the position of the *w*-(0002) reflection, which would systematically hide these grains in the respective dark-field images.

Further studies will concentrate on the structure at the onset of wurtzite-rocksalt phase transition region, and the direct observation of the polarization inversion itself by convergent beam electron diffraction (CBED). In addition to dark-field imaging in specific diffraction conditions [52] , the CBED method was proven to be able to identify the polarization direction in wurtzite GaN [53] and AlN crystals [40,54].

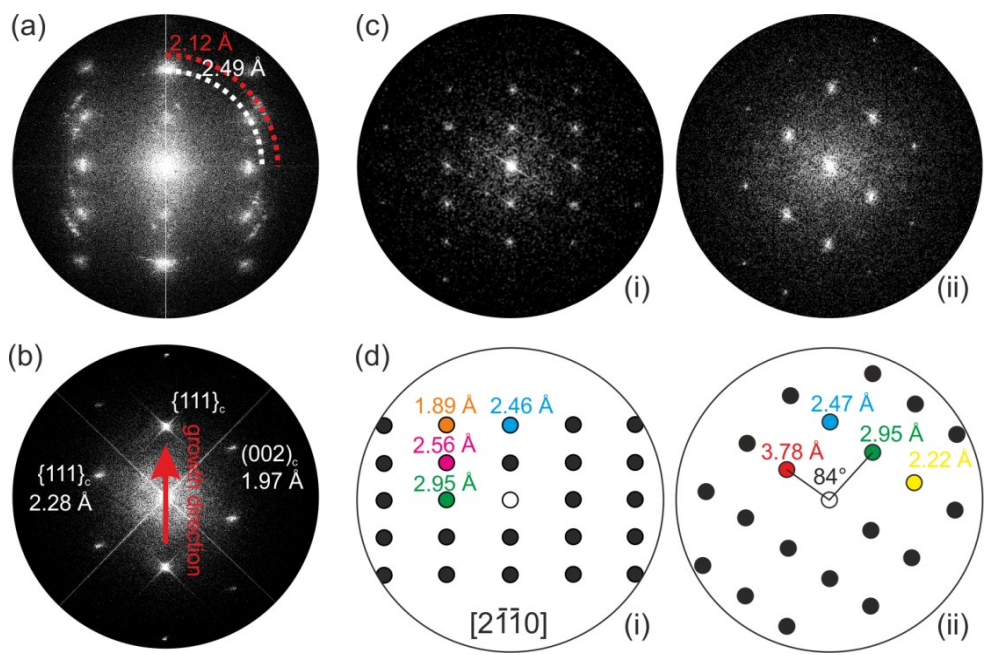

**Figure S6:** FFT study from HRTEM micrographs on the structural variations present in the AlScN film. (a) Intensities with polycrystalline character appear close to the bottom electrode. Intensities along the red circle line might indicate the formation of nano-precipitates of a cubic phase with lattice parameters different to the modified wurtzite-type structure. (b) FFT of a single grain showing a cubic [110] pattern and interplanar distances matching with a high-pressure AlN phase from literature. (c) FFTs from two neighbouring grains and (d) a graphical representation: The grain in (i) shows the $[2\bar{1}\bar{1}0]$ orientation of the wurtzite-type structure, whereas (ii) depicts a grain of different orientation but coinciding reflections along the direction of growth according to the matching interplanar distances along *w*-[000l] (d~2.47Å). However, the determined d-values and angles could not be attributed to the wurtzite-type structure of hexagonal AlScN.

A HRTEM image of the Al$_{0.57}$Sc$_{0.43}$N/electrode interface with corresponding FFT patterns is given in **Figure S7**. In spite of the less than perfect crystal structure and morphology of the sample and its general polycrystalline character, c-axis alignment in the oriented parts of the sample can be observed down the first few nanometers of film growth.

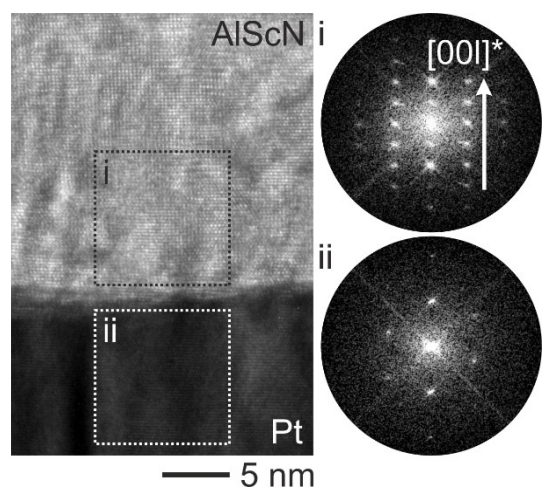

Figure S7: HRTEM image of the Al$_{0.57}$Sc$_{0.43}$N/electrode interface and corresponding FFTs.

Kinematic simulations (Emap Version 1.0, AnaliTEX) of ED pattern demonstrate the intensity variations with increasing internal parameter *u* up to *u*=1/2 c (**Figure S**8), meaning a transition from the polar wurtzite-type into a non-polar configuration and increasing Sc concentration in the alloy. In the extreme cases the parent AlN phase shows identifiable variations in diffracted intensities of type (0,k,-k,l+1) in the depicted $[2\bar{1}\bar{1}0]$ pattern. However, the incorporation of Sc atoms on Al lattice positions increases electron scattering, giving rise to higher intensity (with Sc concentration) and a less pronounced intensity variation with increasing *u* on these lattice positions. On the onset of the proposed transition from the polar to non-polar phase (at x = 0.43) we demonstrate, that the intensity variations are hardly differentiable in the simulations even for low indexed reflections, e.g. $(01\bar{1}1)$. Therefore we reason that a reliable identification of polar and non-polar configurations of the wurtzite-type structure is not feasible using conventional and precession electron diffraction techniques.

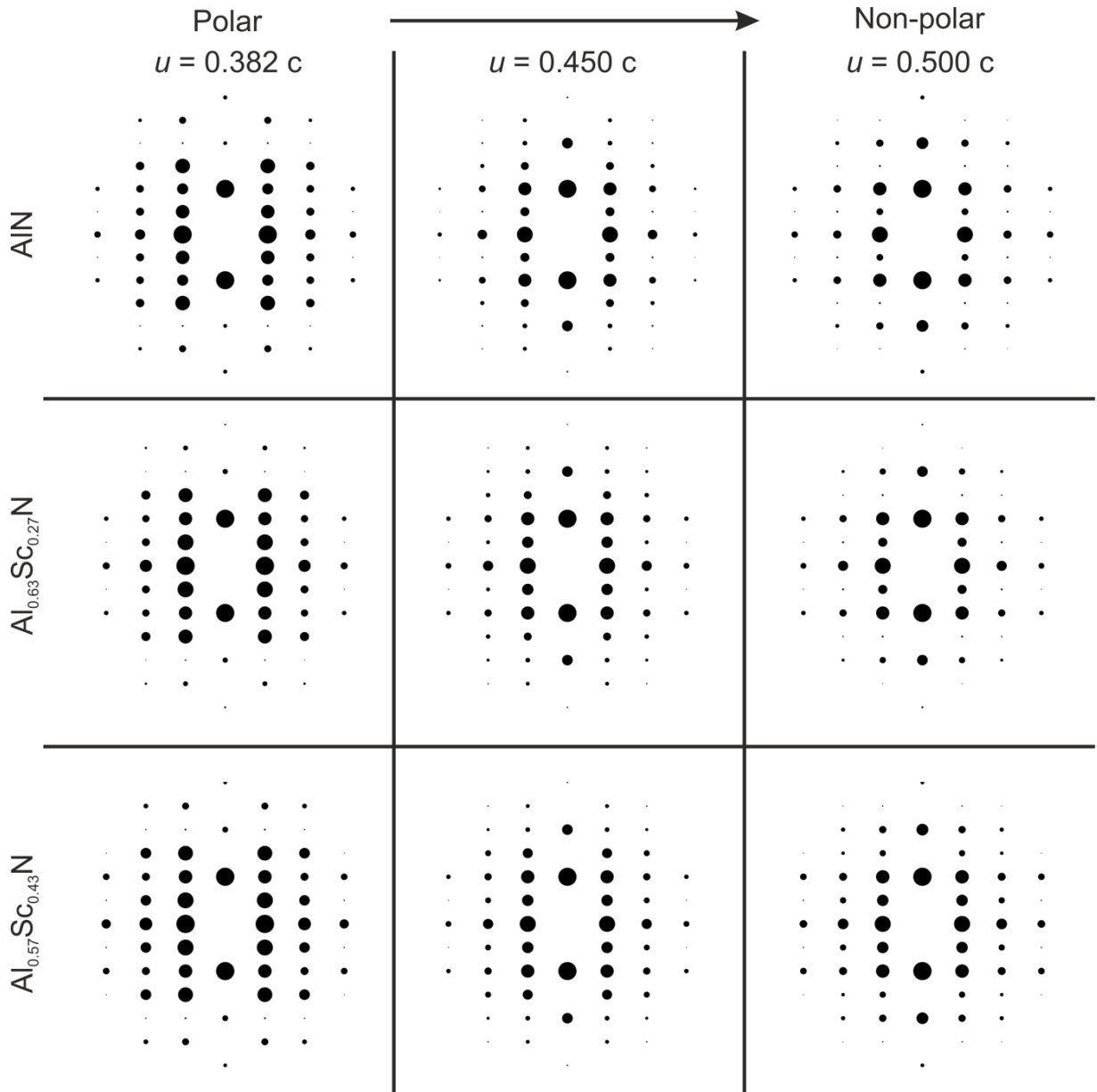

Figure S8: Kinematic simulation of electron diffraction pattern investigating the factors of Sc concentration and the internal parameter $u$ on the reflection intensities.